\documentclass[aps,prb,floatfix,twocolumn,showpacs]{revtex4}%
\usepackage{amsmath}
\usepackage{graphicx}
\usepackage{amsfonts}
\usepackage{amssymb}%
\providecommand{\U}[1]{\protect\rule{.1in}{.1in}}
\providecommand{\U}[1]{\protect\rule{.1in}{.1in}}
\providecommand{\U}[1]{\protect\rule{.1in}{.1in}}

\begin{document}
\title{Gate-dependent spin-torque in a nanoconductor-based spin-valve}
\author{Audrey Cottet}
\affiliation{Laboratoire Pierre Aigrain, Ecole Normale Sup\'{e}rieure, CNRS (UMR 8551),
Universit\'{e} P. et M. Curie, Universit\'{e} D. Diderot, 24 rue Lhomond,
75231 Paris Cedex 05, France}
\date{\today}
\pacs{73.23.-b, 75.75.+a, 85.75.-d}

\begin{abstract}
This article discusses the spin-torque effect in a spin-valve made out of two
ferromagnetic leads connected through a coherent nanoconductor (NC), in the
limit where a single channel of the NC lies near the Fermi energy of the
leads. Due to quantum interferences inside the NC, the spin-torque presents
clear qualitative differences with respect to the case of a multichannel
disordered spin-valve. In particular, it can be modulated with the NC gate
voltage. In principle, this modulation can be observed experimentally,
assuming that the spin-torque affects a ferromagnetic nano-domain in direct
contact with the NC.

\end{abstract}
\maketitle

\section{Introduction}

The study of spin-dependent transport in ferromagnetic hybrid structures has
raised an intense activity in the context of the development of spin
electronics, or spintronics\cite{Prinz}. The most simple and illustrative
spintronics device is the spin-valve. It consists of two ferromagnetic layers
separated by a non-magnetic spacer, which can be conducting or insulating. The
charge current through a spin-valve depends on the relative orientation of the
ferromagnets' magnetizations. This so-called magnetoresistance effect has
allowed the development of new kinds of field-sensing and magnetic memory
devices\cite{Baibich,Binasch}. Conversely, the relative orientation of the
ferromagnets' magnetizations can be modified by a spin-torque
effect\cite{Slonczewski}, which corresponds to an absorption of spin-currents
by the ferromagnets. When the spin-valve spacer is a multichannel disordered
metal, the spin-torque appears only at finite bias\cite{Waintal}. In the case
of a thin ballistic spacer, a torque due to an indirect exchange coupling
between the ferromagnets can also appear in equilibrium conditions, due to a
RKKY-like
interaction\cite{Baibich,grundbergRKKY,parkin,spindepRKKY,spindepRKKYscat}.
The theoretical description of transport in spin-valves is now well developed,
in both the multichannel ballistic and multichannel diffusive regimes (see
e.g. Refs.~\onlinecite{Stiles,Schep,Haney} and \onlinecite{Valet,BrataasReview,Rychkov,Waintal}).

Recently, a gate-controlled magnetoresistance effect has been observed in
spin-valves based on coherent few-channels nanoconductors such as carbon
nanotubes\cite{Sahoo,Man,CottetSST} or self-assembled InAs quantum
dot\cite{Hamaya1,Hamaya2}, placed at low temperatures. The portion of
nanoconductor between the two contacts is subject to a strong electronic
confinement, which leads to the existence of resonant states whose energy can
be shifted by using an electrostatic gate. This allows a strong
gate-modulation of the conductance and magnetoresistance through the device.
However, the spin-torque effect in this kind of device has raised little
attention so far\cite{Romeo,Mu}. This paper discusses the spin-torque effect
in the case where the spin-valve spacer is a coherent nanoconductor (NC) with
a single channel near the Fermi energy of the leads. A non-interacting
scattering formalism is used. The torque felt by each ferromagnet varies with
the NC gate voltage. The spin-activity of the NC/ferromagnet interfaces stems
from the spin dependence of interfacial transmission probabilities and from
the Spin-Dependence of Interfacial scattering Phase Shifts (SDIPS). I first
discuss analytically various limits, in order to emphasize the role of the
different parameters and the qualitative differences with the case of
multichannel disordered spacers. In the latter case, a finite SDIPS is
necessary to obtain an out-of-plane torque component, due to an SDIPS-induced
interfacial effective
field\cite{Stiles,BrataasReview,Brataas1,Brataas2,Dani,Xia}. However, in the
coherent case, this effect can generally not be disentangled from the indirect
exchange coupling between the two ferromagnets, which also gives an
out-of-plane contribution to the torque. Another striking result is that a
Slonczewski in-plane torque can occur even in the limit of spin-independent
interfacial transmission probabilities, due to quantum interferences inside
the NC, which lead to a SDIPS-induced spin-filtering effect. In the
multichannel case, the out-of-plane torque is usually expected to be much
smaller than the in-plane torque, because due to fluctuations of the SDIPS
from one channel to another, the out-of-plane torque almost averages
out\cite{Stiles,BrataasReview}. In contrast, in a NC-based spin-valve, it is
sometimes possible to choose whether the out-of-plane non-equilibrium
contribution to the torque is larger or smaller than the in-plane
contribution, just by changing the NC gate voltage. I finally discuss the
measurability of the spin-torque effect in a NC-based spin-valve. It is
necessary to assume that the spin-torque affects a nanodomain in direct
contact with the NC. This domain can belong to a wider ferromagnetic contact,
similarly to what is observed for torque experiments realized with quantum
point contacts\cite{Chen}.

Note that the non-interacting scattering model used in this article can be
experimentally relevant in the case where the contacts between the
nanoconductor and the ferromagnets have a sufficiently high capacitance. This
was clearly the case for instance in Ref. \cite{Man}, which presents
magnetoresistance data for a spin valve made out of a Single-Wall carbon
Nanotube with PdNi contacts. The conductance of the device versus bias voltage
and gate voltage clearly indicates the absence of interaction effects such as
Coulomb blockade. Therefore, the gate variations of the conductance and
magnetoresistance through the device could be well interpreted using a
non-interacting scattering model similar to the one discussed in the present
article. In the case of contacts with a smaller capacitance, one should use an
interacting description, based for instance on an Anderson-like hamiltonian.

\section{Theoretical model\label{model}}

\begin{figure}[ptb]
\includegraphics[width=1.\linewidth]{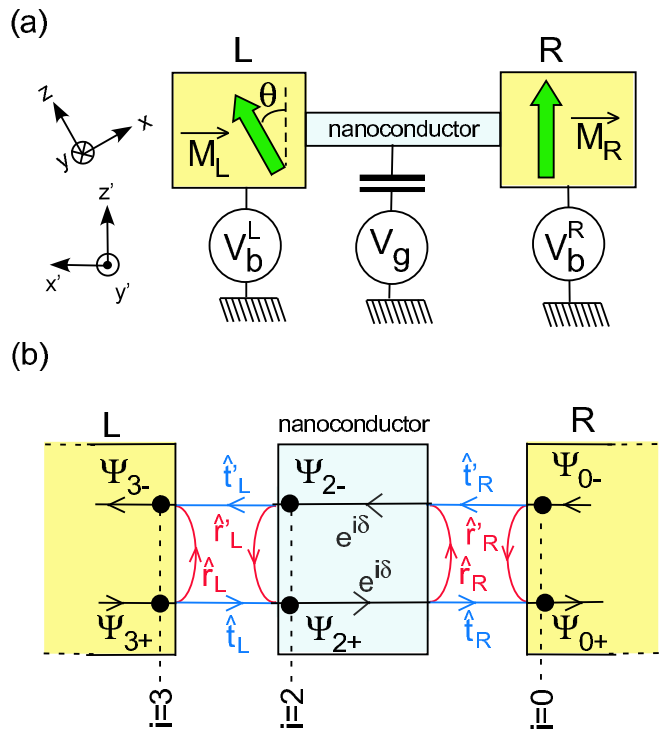}\label{device2}\caption{Panel
a: Scheme of the spin-valve device considered in this article. A ballistic
nanoconductor is connected to two ferromagnets $L$ and $R$ with magnetizations
$\vec{M}_{L}$ and $\vec{M}_{R}$ which form an angle $\theta$. The
nanoconductor is capacitively coupled to a gate biased with a voltage $V_{g}$.
Panel b: Scattering model of the device for channel $m$. The interface between
the ferromagnet $Q\in\{L,R\}$ and the nanoconductor transmits and reflects
electrons with amplitudes $\hat{t}_{Q}$,$\hat{t}_{Q}^{\prime}$ and $\hat
{r}_{Q}$, $\hat{r}_{Q}^{\prime}$ which have a $2\times2$ structure in spin
space (see text). The electrons acquire a winding phase $\delta$ while they
cross the nanoconductors. The rightgoing/leftgoing wavefunction $\Psi_{i\pm}$
at spot $i\in\{0,2,3\}$ is related to the incoming wavefunctions $\Psi_{3+}$
and $\Psi_{0-}$ by $\Psi_{i\pm}=\Gamma_{i\pm}\Psi_{0-}+\Lambda_{i\pm}\Psi
_{3+}$. In the text, we express the spin torque on $L$ in terms of
$\Lambda_{2\pm}$ and $\Gamma_{2\pm}$, and the conductance through the device
in terms of $\Lambda_{0+}$. In section \ref{col}, we assume that $\vec{M}_{R}$
is fixed while $\vec{M}_{L}$ can move. The spin referential $\{x,y,z\}$ is
such that $\vec{M}_{L}=\vec{z}$ while the referential $\{x^{\prime},y^{\prime
},z^{\prime}\}$ is fixed with $\vec{M}_{R}=\vec{z}^{\prime}$.}%
\label{device}%
\end{figure}

I consider a spin-valve made out of a NC with length $\ell$ contacted to two
left and right ferromagnetic electrodes $L$ and $R$ (see Fig.~\ref{device}.a).
The magnetizations of $L$ and $R$ are noted $\overrightarrow{M_{L}}$ and
$\overrightarrow{M_{R}}$. The NC chemical potential can be tuned thanks to a
capacitive gate biased with a voltage $V_{g}$. The dynamics of
$\overrightarrow{M_{L}}$ is affected by a spin-transfer torque $\vec{T}$ which
is due to the spin-dependent scattering of electrons by $L$ and $R$. When
$\overrightarrow{M_{L}}$ and $\overrightarrow{M_{R}}$ \ are non-colinear, the
spin-current incident on $L$ can have components perpendicular to
$\overrightarrow{M_{L}}$, which are not conserved across $L$. However, the
total momentum of the circuit must be conserved. The torque $\vec{T}$
corresponds to an absorption of the non-conserved spin-currents by
$\overrightarrow{M_{L}}$ \cite{Slonczewski}. In real samples, the electronic
transport inside $L$ leads to a relaxation of spin colinearly to
$\overrightarrow{M_{L}}$ (transverse spin-dephasing) because spins components
parallel and antiparallel to $\overrightarrow{M_{L}}$ quickly loose their
coherence with respect to each other when electrons propagate into
$L$\cite{Stiles,Brataas1,Brataas2}. This occurs on a scale $\xi_{F}$ called
the magnetic-coherence length or transverse spin-dephasing length, which is
typically of the order of a nanometer for a ferromagnetic material like
$\mathrm{Ni}$\cite{Petrashov,Robinson,Blum,Shelukhin}. Hence, if the length of
$L$ exceeds a few nanometers along the transport direction, $\overrightarrow
{M_{L}}$ fully absorbs the perpendicular spin-current transmitted into $L$. In
this case, the torque $\vec{T}$ on electrode $L$ corresponds directly to the
transverse component of the spin-current $\vec{I}_{spin,2}$ just at the right
of $L$, i.e. $\vec{T}=-\vec{I}_{spin,2}+(\vec{I}_{spin,2}.\overrightarrow
{M_{L}})\overrightarrow{M_{L}}/M_{L}$\cite{note}. In this picture, it is
possible to treat the ferromagnet $Q\in\{L,R\}$ as a fermionic reservoir, i.e.
the states with energy $E$ inside $Q$ are populated according to a Fermi
distribution $f_{Q}(E)=1/(1+\exp[(E-E_{F}+eV_{b}^{Q})/k_{B}T])$ with
$V_{b}^{Q}$ the bias voltage applied to $Q$, and $E_{F}$ the Fermi energy of
the leads\cite{e}.

In this paper, the electronic transport inside the NC is described with the
Landauer-B\"{u}ttiker scattering formalism\cite{Blanter}. The ferromagnetic
nature of contact $Q$ is taken into account through the spin-dependence of the
electronic scattering matrix $\tilde{S}_{Q}$ between the ferromagnet $Q$ and
the NC, and the transverse spin-dephasing hypothesis inside $Q$. One important
specificity of NCs is the strong energy separation between transverse modes.
As a result, one can reach a regime where a single mode $m$ of the NC lies
near the Fermi energy of the reservoirs (i.e., at a distance smaller than
$k_{B}T$ or the level width). The purpose of this work is to study how the
contribution of mode $m$ to the torque $\vec{T}$ evolves with the leads bias
voltages $V_{b}^{L(R)}$ and the NC gate voltage $V_{g}$. From section
\ref{model} to \ref{gate}, the different spin components are given in a
referential $\{x,y,z\}$ attached to $\overrightarrow{M_{L}}=\vec{z}$ (see
Fig.\ref{device}.a). The matrices $\tilde{S}_{L}$ and $\tilde{S}_{R}$ can be
expressed as $\tilde{S}_{L}=S_{L}$ and $\tilde{S}_{R}=U(\theta)S_{R}%
U^{-1}(\theta)$, with, in the scattering space\cite{Blanter},
\begin{equation}
S_{Q}=\left[
\begin{array}
[c]{cc}%
\hat{r}_{Q} & \hat{t}_{Q}^{\prime}\\
\hat{t}_{Q} & \hat{r}_{Q}^{\prime}%
\end{array}
\right]  \label{Sltild}%
\end{equation}
for $Q\in\{L,R\}$,%
\begin{equation}
U(\theta)=\left[
\begin{array}
[c]{cc}%
\cos\left(  \frac{\theta}{2}\right)  \hat{\sigma}_{0}-i\sin\left(
\frac{\theta}{2}\right)  \hat{\sigma}_{y} & 0\\
0 & \cos\left(  \frac{\theta}{2}\right)  \hat{\sigma}_{0}-i\sin\left(
\frac{\theta}{2}\right)  \hat{\sigma}_{y}%
\end{array}
\right]
\end{equation}
I note $\hat{\sigma}_{x}$, $\hat{\sigma}_{y}$ and $\hat{\sigma}_{z}$ the Pauli
matrices in spin space and $\hat{\sigma}_{0}$ the identity matrix in spin
space. The reflection and transmission matrices between the ferromagnet $Q$
and the NC, noted $\hat{r}_{Q}$, $\hat{r}_{Q}^{\prime}$ and $\hat{t}_{Q}%
,\hat{t}_{Q}^{\prime}$ respectively, are defined in Fig. \ref{device}.b. These
matrices have a structure in spin space, i.e.,
\begin{equation}
\hat{r}_{Q}=\left[
\begin{array}
[c]{cc}%
\sqrt{1-T_{Q}^{u}}e^{i\varphi_{Q}^{u}} & 0\\
0 & \sqrt{1-T_{Q}^{d}}e^{i\varphi_{Q}^{d}}%
\end{array}
\right]  \label{r}%
\end{equation}%
\begin{equation}
\hat{r}_{Q}^{\prime}=\left[
\begin{array}
[c]{cc}%
\sqrt{1-T_{Q}^{u}}e^{i\bar{\varphi}_{Q}^{u}} & 0\\
0 & \sqrt{1-T_{Q}^{d}}e^{i\bar{\varphi}_{Q}^{u}}%
\end{array}
\right]  \label{rp}%
\end{equation}
and%
\begin{equation}
\hat{t}_{Q}=\hat{t}_{Q}^{\prime}=\left[
\begin{array}
[c]{cc}%
i\sqrt{T_{Q}^{u}}e^{i\frac{\varphi_{Q}^{u}+\bar{\varphi}_{Q}^{u}}{2}} & 0\\
0 & i\sqrt{T_{Q}^{d}}e^{i\frac{\varphi_{Q}^{d}+\bar{\varphi}_{Q}^{d}}{2}}%
\end{array}
\right]  \label{t}%
\end{equation}
The number of parameters occurring in $S_{Q}$ has been minimized by assuming
flux conservation and spin conservation along $\overrightarrow{M_{Q}}$ by the
$Q$/NC interface. The $u$ and $d$ indices refer to majority and minority spin
species for each ferromagnet considered. I note $T_{Q}^{u(d)}=1-R_{Q}^{u(d)}$
the transmission probability for a majority(minority) spin across contact $Q$,
while $\varphi_{Q}^{u(d)}$ and $\bar{\varphi}_{Q}^{u(d)}$ are the reflection
phases for majority(minority) spins on the left and right side of the $Q$/NC
interface respectively. The values of the interfacial transmission phases are
imposed by those of the reflection phases, which explains the shape of the
phase factors in Eq.(\ref{t}) [see Ref. \onlinecite{BCIGF} for details]. Note
that the values of the interface parameters $T_{Q}^{u(d)}$, $\varphi
_{Q}^{u(d)}$ and $\bar{\varphi}_{Q}^{u(d)}$ are difficult to predict since
they can depend on the microscopic details of the ferromagnet/NC contacts.
However, they can be considered as fitting parameters which have to be
determined for each sample. Such an approach was already used successfully to
interpret quantitatively spin-dependent transport experiments in spin valves
and multiterminal circuits based on single wall carbon
nanotubes\cite{Sahoo,Man,CPF}. Electrons acquire a winding phase $\delta$
while crossing the NC. This phase can be tuned with the NC gate voltage
$V_{g}$. It also depends on the electronic energy $E$ (see Section \ref{col}).

The conductance and magnetoresistance corresponding to the above model have
already been studied theoretically in Ref.~\onlinecite{cottet06}. Due to
quantum interferences inside the NC, these signals depend on $\delta$ and thus
on $V_{g}$. Reference~\onlinecite{Romeo} has discussed the torque in a
one-dimensional spin-valve model based on a single channel
Blonder-Tinkham-Klapwijk approach\cite{BTK}. This case is very different from
the one discussed in the present paper, since in Ref.~\onlinecite{Romeo} the
whole ferromagnetic contacts are modeled as delta-function potential barriers
which produce no transverse spin-dephasing. Reference~\onlinecite{Mu} has used
an Anderson-hamiltonian approach. However, these authors have studied only the
in-plane out-of-equilibrium torque and they did not take into account the
SDIPS. In the present approach, the ferromagnet/NC interfaces could be
alternatively modeled as delta-function potential barriers. However, this
would impose a given relation between the interfacial transmissions and
scattering phases. Equations (\ref{r}-\ref{t}) are more general since they can
account any type of interface potential profile. They also allow to study
separately the effects of the spin-dependence of the interface transmission
probabilities and of the SDIPS. It has been shown that these two properties
affect the device conductance $G$ in qualitatively different
ways\cite{cottet06}. Qualitative differences are also expected for the spin-torque.

\section{General expression of the spin-torque\label{generalExpr}}

For simplicity, one can assume that, at any energy, mode $m$ is not coupled to
the other modes of the NC upon scattering by the NC/ferromagnet contacts. In
this case, the torque acting on the left magnetization writes $\vec{T}=\vec
{T}_{m}+\vec{C}$ with a separate contribution $\vec{T}_{m}$ from mode $m$. The
contribution $\vec{C}$ accounts for other modes which are far from the Fermi
energy of the reservoirs. It is convenient to decompose $\vec{T}_{m}$ as
$\vec{T}_{m}^{eq}+\vec{T}_{m}^{tr}$, with a finite bias contribution $\vec
{T}_{m}^{tr}$ and an equilibrium term $\vec{T}_{m}^{eq}$ which exists in the
absence of a bias voltage, i.e. when $f_{0(3)}(E)=f_{eq}(E)=1/(1+\exp
[(E-E_{F})/k_{B}T])$. The parametrization introduced in section \ref{model}
leads to\cite{note2}
\begin{equation}
\vec{T}_{m}^{eq}=%
%TCIMACRO{\tint }%
%BeginExpansion
{\textstyle\int}
%EndExpansion
dE\left[  A_{23}^{y}(E)+A_{20}^{y}(E)\right]  f_{eq}(E)\vec{y} \label{torque2}%
\end{equation}
and%
\begin{align}
\vec{T}_{m}^{tr}  &  =%
%TCIMACRO{\tint }%
%BeginExpansion
{\textstyle\int}
%EndExpansion
dEA_{20}^{x}(E)\left[  f_{0}(E)-f_{3}(E)\right]  \vec{x}\nonumber\\
&  +%
%TCIMACRO{\tsum \nolimits_{i\in\{0,3\}}}%
%BeginExpansion
{\textstyle\sum\nolimits_{i\in\{0,3\}}}
%EndExpansion%
%TCIMACRO{\tint }%
%BeginExpansion
{\textstyle\int}
%EndExpansion
dEA_{2i}^{y}(E)\left[  f_{i}(E)-f_{eq}(E)\right]  \vec{y} \label{bloup}%
\end{align}
with $A_{20}^{\mu}(E)=Tr_{\sigma}\left[  \sigma_{\mu}\left(  \Gamma_{2-}%
\Gamma_{2-}^{\dag}-\Gamma_{2+}\Gamma_{2+}^{\dag}\right)  \right]  /4\pi$ and
$A_{23}^{\mu}(E)=Tr_{\sigma}\left[  \sigma_{\mu}(\Lambda_{2-}\Lambda
_{2-}^{\dag}-\Lambda_{2+}\Lambda_{2+}^{\dag})\right]  /4\pi$ for $\mu
\in\{x,y\}$. The in-plane and out-of-plane torques correspond to $\vec{x}$ and
$\vec{y}$ components respectively. Here, $\Gamma_{2\mp}$ [$\Lambda_{2\mp}$]
are the coefficients obtained when decomposing the left[right]-going
wavefunction associated to $m$ just on the right of $L$ in terms of the modes
incoming from $L$ and $R$ (see Fig.\ref{device}.b). These coefficients can be
expressed in terms of the parameters introduced in Section \ref{model}. The
trace in the above expressions runs over the spin index $\sigma$. We will see
in section \ref{analytic} that mode $m$ can contribute to the equilibrium
value of the torque ($\vec{T}_{m}^{eq}\neq0$) because one has in the general
case $A_{23}^{y}(E)\neq-A_{20}^{y}(E)$. Since $f_{eq}(E)$ appears in the
integrand of Eq.(\ref{torque2}), in principle, $\vec{T}_{m}^{eq}$ depends on
the properties of channel $m$ on a wide range of energies for which
$T_{Q}^{u(d)}$, $\varphi_{Q}^{u(d)}$, and $\bar{\varphi}_{Q}^{u(d)}$ should be
energy dependent. By analogy, $\vec{C}$ can also be finite although it
accounts for the contribution of modes which are far from the Fermi energy of
the reservoirs. The full values of $\vec{T}_{m}^{eq}$ and $\vec{C}$ depend on
the whole band structure of the NC and ferromagnets. However, if $V_{b}%
^{L(R)}$ and $V_{g}$ are too small to bring other modes than $m$ close to
$E_{F}$, $\vec{C}$ can be considered as independent from the gate and bias
voltages. The main purpose of this work is study the gate and bias dependences
of the torque, which are contained in $\vec{T}_{m}=\vec{T}_{m}^{eq}+\vec
{T}_{m}^{tr}$. Note that when a finite bias voltage $V_{b}$ is applied to the
left reservoir ($V_{b}^{L}=V_{b}$ and $V_{b}^{R}=0$), one obtains, in the low
temperature linear regime $eV_{b}\ll k_{B}T\ll T_{L[R]}^{u(d)}\hbar
v_{F}/2\ell$,
\begin{equation}
\vec{T}_{m}^{tr}=\left.  eV_{b}A_{20}^{x}\vec{x}-eV_{b}A_{23}^{y}\vec
{y}\right\vert _{E=E_{F}} \label{8a}%
\end{equation}
This is not equivalent to applying the bias voltage to the right reservoir
($V_{b}^{L}=0$ and $V_{b}^{R}=-V_{b}$), since one finds in this second case%
\begin{equation}
\vec{T}_{m}^{tr}=\left.  eV_{b}A_{20}^{x}\vec{x}+eV_{b}A_{20}^{y}\vec
{y}\right\vert _{E=E_{F}} \label{8b}%
\end{equation}
One can check that the expressions of $A_{20}^{x}$, $A_{20}^{y}$ and
$A_{23}^{y}$ involve a denominator
\[
D(\theta)=\left\vert \beta_{u{}u}\beta_{d{}d}\cos^{2}(\theta)+\beta_{u{}%
d}\beta_{u{}d}\sin^{2}(\theta)\right\vert ^{2}%
\]
with $\beta_{s{}s^{\prime}}=1-e^{i\phi_{s,s^{\prime}}}\sqrt{R_{L}^{s}%
R_{R}^{s^{\prime}}}$, $\phi_{s,s^{\prime}}=2\delta+\bar{\varphi}_{L}%
^{s}+\varphi_{R}^{s^{\prime}}$ and spin indices $(s,s^{\prime})\in\{u,d\}^{2}$
defined in section \ref{model}. This denominator expresses the fact that
electrons are subject to multiple reflections between the two ferromagnets.
This leads to resonances which appear as peaks in the conductance
$G=(e^{2}/\hbar)Tr_{\sigma}\left[  \Lambda_{0+}^{\dag}\Lambda_{0+}\right]  $
of the spin-valve versus $\delta$ (see for instance Fig. \ref{Fig2}).
Similarly, the torque can strongly depend on $\delta$, as shown below.

\section{Analytical expressions of the torque in various
limits\label{analytic}}

This section discusses analytically various limiting cases, which are not
necessarily obvious to reach in practice, but allow to understand the role the
different parameters.

\subsection{Case of a spin-independent L/NC contact\label{si}}

I first assume that the scattering matrix $\tilde{S}_{L}$ describing the
contact between the ferromagnet L and the NC is not spin-dependent, i.e.
$T_{L}^{u(d)}=T_{L}=1-R_{L}$, $\varphi_{L}^{u(d)}=\varphi_{L}$, and
$\bar{\varphi}_{L}^{u(d)}=\bar{\varphi}_{L}$. One finds $A_{20}^{y}=A_{23}%
^{y}=0$, thus there is no equilibrium torque and no out-of-plane torque in
this case. In contrast, one finds a finite out-of equilibrium in-plane torque
($\vec{T}_{m}^{tr}\neq0$) since%
\begin{align}
A_{20}^{x}  &  =T_{L}\frac{\sin(\theta)}{4\pi D(\theta)}\left[  (T_{L}%
-2)(T_{R}^{d}-T_{R}^{u})\right. \nonumber\\
&  \left.  +2\sqrt{R_{L}}\left(  \sqrt{R_{R}^{d}}T_{R}^{u}\cos[\phi_{d}%
]-\sqrt{R_{R}^{u}}T_{R}^{d}\cos[\phi_{u}]\right)  \right]  \label{sl}%
\end{align}
with $\phi_{u(d)}=2\delta+\bar{\varphi}_{L}+\varphi_{R}^{u(d)}$. This effect
is similar to the spin-filtering torque discussed by
Slonczewski\cite{Slonczewski}. In the present case, the spin filtering is not
due to the interface matrix $\tilde{S}_{L}$, which is spin conserving, but to
$L$ itself, since a transverse spin-dephasing occurs inside $L$.
Interestingly, the torque can be controlled with the NC gate voltage since
$\delta$ occurs in Eq.(\ref{sl}). From the above equation, if $T_{R}^{u}%
=T_{R}^{d}=T_{R}$, the coefficient $A_{20}^{x}$ remains finite, i.e.%
\[
A_{20}^{x}=\sqrt{R_{L}R_{R}}T_{L}T_{R}\left(  \cos[\phi_{u}]-\cos[\phi
_{d}]\right)  \frac{\sin(\theta)}{2\pi D(\theta)}%
\]
The existence of a finite torque may seem surprising in this case. Indeed, if
the right NC/ferromagnet contact was considered alone (semi-infinite
geometry), the current in the NC would not be spin-polarized since $T_{R}%
^{u}=T_{R}^{d}$. However, one should keep in mind that quantum interferences
occur inside the NC. In the presence of a SDIPS at contact R, the whole F/NC/F
device behaves as a spin polarizer along $\overrightarrow{M_{R}}$, because
spins parallel and antiparallel to $\overrightarrow{M_{R}}$ are resonant
inside the NC for different energies\cite{cottet06}. To confirm the crucial
role of quantum interferences in this effect, one can check that $A_{20}^{x}$
vanishes for $T_{L}=1$.

\subsection{Case with no SDIPS\label{noSDIPS}}

I now consider a case where $\tilde{S}_{L}$ and $\tilde{S}_{R}$ are both spin
dependent but there is no SDIPS, i.e. $\varphi_{L[R]}^{u(d)}=\varphi_{L[R]}$
and $\bar{\varphi}_{L[R]}^{u(d)}=\bar{\varphi}_{L[R]}$. In this limit, one can
check%
\begin{align}
A_{20}^{y}  &  =\sin\left[  \theta\right]  \sin\left[  2(\delta+\bar{\varphi
}_{L}+\varphi_{R})\right] \nonumber\\
&  \times\left(  \sqrt{R_{L}^{u}}\left(  T_{L}^{d}-2\right)  -\sqrt{R_{L}^{d}%
}\left(  T_{L}^{u}-2\right)  \right) \nonumber\\
&  \times\left(  T_{R}^{u}\sqrt{R_{R}^{d}}-T_{R}^{d}\sqrt{R_{R}^{u}}\right)
/4\pi\label{bi2}%
\end{align}%
\begin{align}
A_{23}^{y}  &  =\sin\left[  \theta\right]  \sin\left[  2(\delta+\bar{\varphi
}_{L}+\varphi_{R})\right] \nonumber\\
&  \times\left(  \sqrt{R_{R}^{u}}\left(  T_{R}^{d}-2\right)  -\sqrt{R_{R}^{d}%
}\left(  T_{R}^{u}-2\right)  \right) \nonumber\\
&  \times\left(  T_{L}^{u}\sqrt{R_{L}^{d}}-T_{L}^{d}\sqrt{R_{L}^{u}}\right)
/4\pi\label{bi4}%
\end{align}
and, using $T_{L(R)}^{u[d]}=T_{u[d]}$, $\varphi_{L[R]}^{u(d)}=\varphi$ and
$\bar{\varphi}_{L[R]}^{u(d)}=\bar{\varphi}$,
\begin{equation}
A_{20}^{x}=(T_{u}-T_{d})\left(  T_{u}+T_{d}-T_{u}T_{d}\right)  \sin\left[
\theta\right]  \sin^{2}\left[  \phi/2\right]  /2\pi\label{bi3}%
\end{equation}
with $\phi=2\delta+\bar{\varphi}+\varphi$. Interestingly, Eqs. (\ref{bi2}) and
(\ref{bi4}) give $A_{20}^{y}=A_{23}^{y}$ for $T_{L}^{u(d)}=T_{R}^{u(d)}$.
Therefore, one can obtain an out-of-plane contribution to the equilibrium
torque without a SDIPS ($\vec{T}_{m}^{eq}\neq0$), even if the NC/ferromagnet
contacts are symmetric. This effect is a corollary of the\ non-local-exchange
coupling mediated by itinerant-electrons, which has been observed between two
ferromagnets connected through a very thin normal metal
spacer\cite{grundbergRKKY,Baibich,parkin}. The non-local exchange can be
explained in terms of a spin-dependent RKKY interaction, which is naturally
taken into account by scattering
descriptions\cite{spindepRKKYscat,Waintal,Xiao}. Interestingly, this effect
does not occur when the central conductor of the spin-valve is a diffusive
metallic electrode\cite{Stiles}, because it vanishes in the limit of a large
number of channels in the presence of disorder\cite{Waintal,BrataasReview}.
From Eqs.(\ref{bi2}-\ref{bi3}), the out-of-equilibrium torque has both an
in-plane component (Slonczewski-like) and out-of-plane component (related to
the interlayer exchange coupling). In contrast, in the case where the
conductor placed between the two ferromagnets is a multichannel diffusive
conductor, there is no out-of-plane non-equilibrium torque when the SDIPS vanishes.

\subsection{Case of two non-spin filtering contacts\label{nsf}}

I now assume that both contacts have spin-independent transmission
probabilities, but a finite SDIPS, i.e. $T_{L[R]}^{u(d)}=T_{L[R]}=1-R_{L[R]}$.
In this case, one finds%
\[
A_{20}^{x}=T_{L}T_{R}\cos[\frac{\bar{\varphi}_{L}^{u}-\bar{\varphi}_{L}^{d}%
}{2}]\Theta
\]%
\begin{equation}
A_{20}^{y}=(2-T_{L})T_{R}\sin[\frac{\bar{\varphi}_{L}^{u}-\bar{\varphi}%
_{L}^{d}}{2}]\Theta\label{af}%
\end{equation}%
\[
A_{20}^{y}-A_{23}^{y}=2(T_{R}-T_{L})\sin[\frac{\bar{\varphi}_{L}^{u}%
-\bar{\varphi}_{L}^{d}}{2}]\Theta
\]
and%
\begin{equation}
\Theta=\sqrt{R_{L}R_{R}}\sin[\frac{\varphi_{R}^{d}-\varphi_{R}^{u}}{2}%
]\sin[\frac{\phi_{u,u}+\phi_{d,d}}{2}]\frac{\sin(\theta)}{\pi D(\theta)}
\label{af2}%
\end{equation}
Thus, there exists out-of-plane contributions to both $\vec{T}_{m}^{eq}$ and
$\vec{T}_{m}^{tr}$. In the multichannel incoherent case, it has already been
found that the SDIPS can cause an out-of-plane torque, proportionally to the
imaginary part of the so-called "mixing conductance". To understand this
effect, one must note that an electron scattered by contact $L$ with a spin
non-colinear to $\overrightarrow{M_{L}}$ precesses around $\overrightarrow
{M_{L}}$ due to $\bar{\varphi}_{L}^{u}\neq\bar{\varphi}_{L}^{d}$. In other
words, the SDIPS causes an effective Zeeman interfacial field along
$\overrightarrow{M_{L}}$. Due to momentum conservation, the electronic
precession around this field leads to an out-of-plane torque on $L$. However,
this picture is more delicate to use in the present case where the existence
of an out-of-plane torque $\vec{T}_{m}^{tr}$ cannot be disentangled from
interference effects, because it requires $T_{L}<1$. Therefore, the
Slonczewski in-plane torque and the out-of-plane indirect exchange effect
discussed above also occur.

\subsection{Case of a left perfectly transmitting contact}

To suppress quantum interferences inside the NC, one can consider the case
$T_{L}^{u(d)}=$ $1$. This gives quite generally $A_{20}^{y}=A_{23}^{y}=0$ and
\[
A_{20}^{x}=(T_{R}^{u}-T_{R}^{d})\sin(\theta)/4\pi
\]
Due to the absence of quantum interferences, interfacial reflection phases are
not relevant anymore. The contribution of mode $m$ to the torque is purely
in-plane. The absence of an out-of-plane torque contribution may seem
surprising since the transmission phases from the NC to $L$ can depend on spin
[see Eq.~(\ref{t})]. However, this property has no physical consequence in the
present model, because of the transverse spin-dephasing occurring in $L$.

\subsection{Case of a left perfectly reflecting contact}

In contrast, the present model gives a purely out-of-plane torque contribution
in the limit $T_{L}^{u(d)}=0$, which leads quite generally to $A_{20}%
^{x}=A_{23}^{y}=0$ and
\begin{align*}
A_{20}^{y}  &  =\frac{\sin(\theta)}{\pi D(\theta)}\left(  (T_{R}^{u}-T_{R}%
^{d})\cos\left[  \frac{\bar{\varphi}_{L}^{u}-\bar{\varphi}_{L}^{d}}{2}\right]
\right. \\
&  -\sqrt{R_{R}^{d}}T_{R}^{u}\cos\left[  \frac{\phi_{u,d}+\phi_{d,d}}%
{2}\right] \\
&  \left.  +\sqrt{R_{R}^{u}}T_{R}^{d}\cos\left[  \frac{\phi_{u,u}+\phi_{d,u}%
}{2}\right]  \right)  \sin[\frac{\bar{\varphi}_{L}^{u}-\bar{\varphi}_{L}^{d}%
}{2}]
\end{align*}
There is no in-plane torque contribution in this case because electrons cannot
cross the $L$/NC interface, and therefore, the spin-filtering effect
considered by Slonczewski is not relevant anymore. One finds $A_{23}^{y}=0$
because electrons can enter the device through the right contact only. The
torque depends on $V_{b}^{R}$ although there is no charge transport. This
counterintuitive result can be understood by noting that since $T_{L}%
^{u(d)}=0$, electrons inside the NC remain in equilibrium with the right
ferromagnet. In this case, changing $V_{b}^{R}$ instead of $V_{g}$ just gives
another way to observe the variations of an equilibrium torque. The
SDIPS-induced interface exchange field at the left contact plays a crucial
role in the establishment of this torque since $A_{20}^{y}=0$ for
$\bar{\varphi}_{L}^{u}=\bar{\varphi}_{L}^{d}$.

\section{Angular and gate dependence of the out-of equilibrium part of the
torque\label{gate}}

As already explained in section \ref{generalExpr}, the absolute value of the
torque felt by $L$ depends on the whole band structure of the NC and
ferromagnets. The purpose of this article is not to calculate this value, but
the gate and bias dependences of the torque, for small applied voltages. If
$V_{b}^{L(R)}$ and $V_{g}$ are too small to bring other modes than $m$ close
to $E_{F}$, they can modify significantly the torque contribution $\vec{T}%
_{m}$ from mode $m$ only. Using constant values for $T_{Q}^{u(d)}$,
$\varphi_{Q}^{u(d)}$ and $\bar{\varphi}_{Q}^{u(d)}$ is a reasonable assumption
in this context. In practice, one can check, using realistic parameters, that
the variations of $\vec{T}_{m}^{eq}$ with $V_{g}$ are likely to be small (see
section \ref{col}). Therefore, I have chosen to focus on the angular and gate
dependence of $\vec{T}_{m}^{tr}$. In this section and the following, I use
$T_{L[R]}^{u(d)}=T_{L[R]}(1\pm P_{L[R]})$. From the previous section, one can
check that the only physically relevant phases are the reflection phases
inside the NC, i.e. $\bar{\varphi}_{L}^{u(d)}$ and $\varphi_{R}^{u(d)}$. I use
below $\varphi_{L(R)}^{u(d)}=\bar{\varphi}_{L(R)}^{u(d)}=\varphi^{u(d)}$.

\begin{figure}[ptb]
\includegraphics[width=0.9\linewidth]{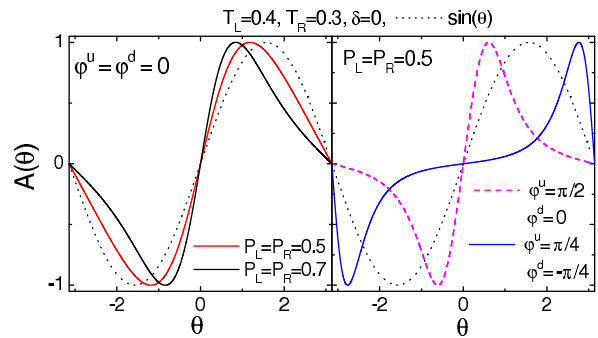}\caption{Function $A(\theta)$
giving the angular dependence of $\vec{T}_{m}^{tr}$ in the low-temperature and
linear-bias limit. The left panel shows the effect of a variation in the
spin-polarization $P_{L[R]}$ of the interface tunnel probabilities, in the
absence of a SDIPS ($\varphi^{u}=\varphi^{d}$). The right panel shows the
effect of a variation in the spin averaged reflection phase $(\varphi
^{u}+\varphi^{d})/2$, for a finite and constant SDIPS $\varphi^{u}-\varphi
^{d}=\pi/2$. For comparison, the function $\sin[\theta]$ is shown with dotted
lines in both panels.}%
\label{Fig1}%
\end{figure}

For simplicity, I discuss the angular dependence of $\vec{T}_{m}^{tr}$ in the
low temperature linear regime [see Eqs. (\ref{8a}) and (\ref{8b})]. This
dependence can be characterized with the function $A(\theta)=(\sin
(\theta)/D(\theta))/\max_{\theta}(\sin(\theta)/D(\theta))$. Indeed, in the
different cases considered analytically in section \ref{analytic}, the
coefficients $A_{20}^{x}$, $A_{20}^{y}$ and $A_{23}^{y}$ are proportional to
$\sin(\theta)/D(\theta)$. I have checked analytically that this property
remains true even if no particular hypotheses are made on $\tilde{S}_{L}$ and
$\tilde{S}_{R}$. The spin-torque is often non-sinusoidal in the multichannel
disordered case (see e.g. Ref.~\onlinecite{Waintal}). In the present case,
when $T_{L}^{u(d)}$ and $T_{R}^{u(d)}$ are close to 1, one finds
$D(\theta)\rightarrow1$, so that $A(\theta)=\sin(\theta)$ to a good
approximation. In the absence of a SDIPS and for $T_{L}^{u(d)}$ and
$T_{R}^{u(d)}$ close to 0, one finds $D(\theta)\rightarrow(1-e^{i\phi}){}^{2}$
with $\phi_{s,s^{\prime}}=\phi$, so that $A(\theta)=\sin(\theta)$ again. In
order to have a non-sinusoidal $A(\theta)$ for small values of $T_{L[R]}%
^{u(d)}$, a finite SDIPS must be used. However, it is also possible to have a
non-sinusoidal $A(\theta)$ for a vanishing SDIPS by using intermediate values
for $T_{L[R]}^{u(d)}$. The left panel of Fig.\ref{Fig1} shows $A(\theta)$ in
the absence of a SDIPS. An increase in $P_{L[R]}$ can help to increase the
anharmonicity of $A(\theta)$. From the right panel of Fig.\ref{Fig1}, it
nevertheless seems that a strong SDIPS more easily leads to a strongly
non-sinusoidal $A(\theta)$. The anharmonicity of $A(\theta)$ can also be
changed with $\delta$ (not shown) and with the value of the spin averaged
reflection phases $(\varphi_{L(R)}^{u}+\varphi_{L(R)}^{d})/2$ (see right panel
of Fig.\ref{Fig1}).

\begin{figure}[ptb]
\includegraphics[width=0.7\linewidth]{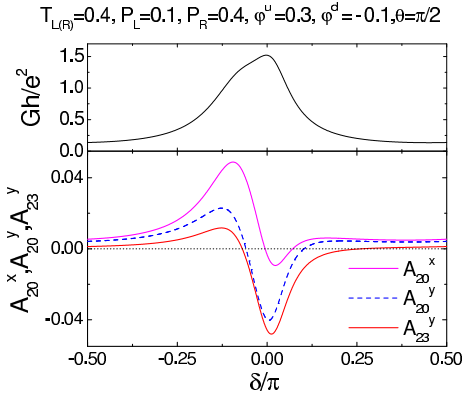}\caption{Conductance $G$ (top
panel) and coefficients $A_{20}^{x}$, $A_{20}^{y}$ and $A_{23}^{y}$
determining the low-temperature linear-bias limit of $\vec{T}_{m}^{tr}$
(bottom panel), as a function of the winding phase $\delta$ through the NC.}%
\label{Fig2}%
\end{figure}

\begin{figure}[ptbptb]
\includegraphics[width=0.5\linewidth]{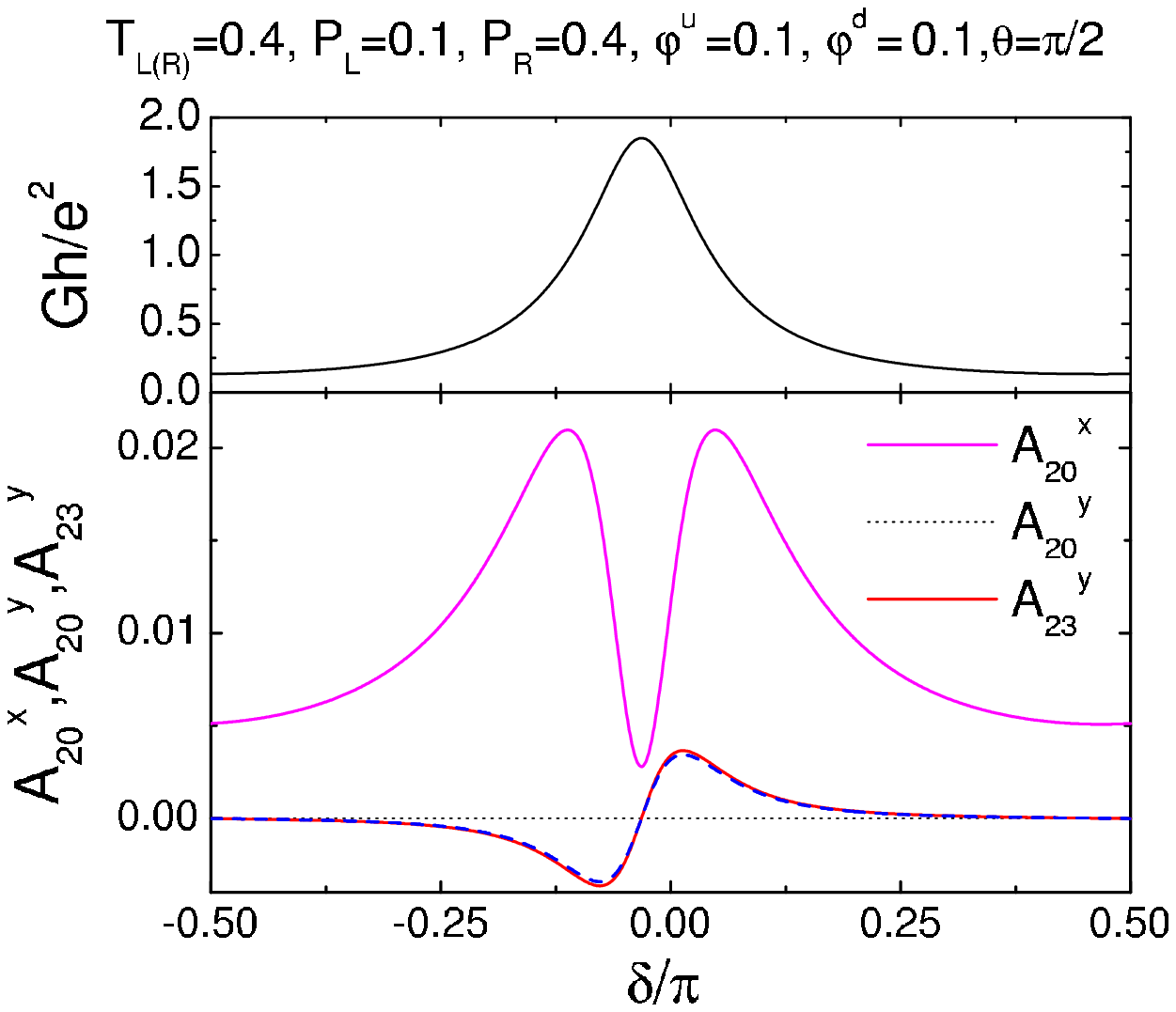}\includegraphics[width=0.5\linewidth]{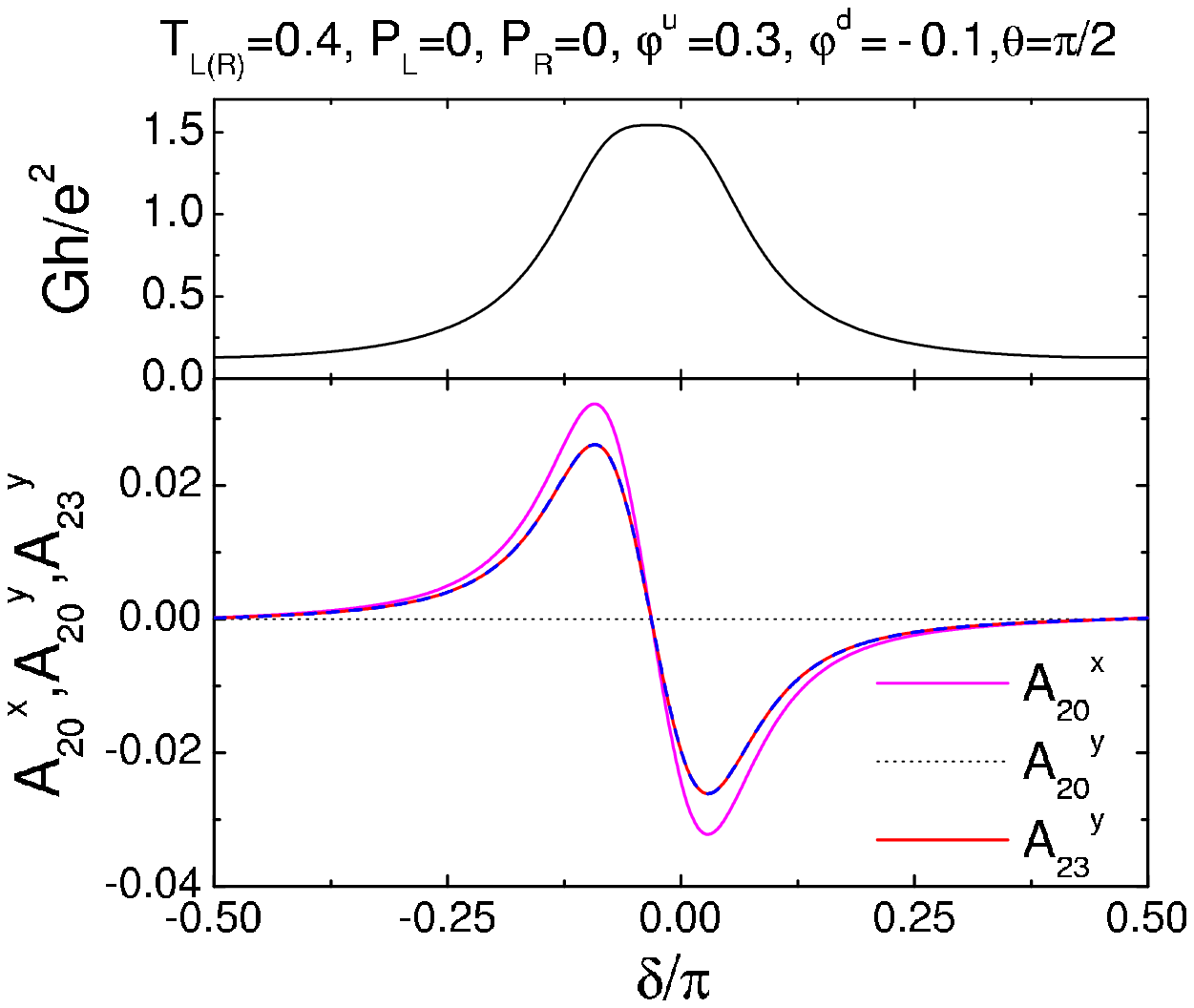}\caption{Conductance
$G$ and coefficients $A_{20}^{x}$, $A_{20}^{y}$ and $A_{23}^{y}$, as a
function of $\delta$. The parameters used here are the same as in
Fig.~\ref{Fig2}, except $\varphi^{u}=\varphi^{d}$ (no SDIPS) in the left
panels, and $P_{L(R)}=0$ in the right panels.}%
\label{Fig5}%
\end{figure}

I now discuss the $\delta$-dependence of $\vec{T}_{m}^{tr}$, in the low
temperature linear regime. This dependence is given by the coefficients
$A_{20}^{x}$, $A_{20}^{y}$ and $A_{23}^{y}$, which are shown in Figure
\ref{Fig2}, for a particular set of interface parameters. These coefficients
are $\pi$-periodic with $\delta$. Remarkably, they can change sign with
$\delta$. In the multichannel  case, the out-of-plane torque is usually
expected to be much smaller than the in-plane torque, because due to
fluctuations of the SDIPS from one channel to another, the out-of-plane torque
almost averages out\cite{Stiles,BrataasReview,Xia}. However, in the present
case, $\left\vert A_{20}^{y}\right\vert $ and $\left\vert A_{23}%
^{y}\right\vert $ can be smaller or larger than $\left\vert A_{20}%
^{x}\right\vert $ depending on the value of $\delta$ considered (see bottom
panel of Fig.\ref{Fig2}). To illustrate the effects of the SDIPS and of the
polarization $P_{L(R)}$ of the interfacial tunnel probabilities, Figure
\ref{Fig5} shows the coefficients $A_{20}^{x}$, $A_{20}^{y}$ and $A_{23}^{y}$
for the same parameters as in Figure \ref{Fig2}, but with $\varphi^{u}%
=\varphi^{d}$ (no SDIPS) in the left panel and $P_{L(R)}=0$ in the right
panel. In the right panel, one has exactly $A_{20}^{y}=A_{23}^{y}$ due to
$T_{L}=T_{R}$ (see section \ref{nsf}). In the left panel, $A_{20}^{y}$ and
$A_{23}^{y}$ are not exactly equal, but the difference is too small to be
visible on the scale of the figure, because the spin-dependent transmission
probabilities $T_{L}(1\pm P_{L})$ are relatively close to $T_{R}(1\pm P_{R})$
(see section \ref{noSDIPS}). In general, using a finite SDIPS allows one to
increase strongly the amplitude of the torque variations, because the SDIPS
tends to induce a spin-splitting of electronics resonances inside the NC,
which strongly spin-polarizes the current through the NC (compare
Fig.~\ref{Fig5},left with Fig.~\ref{Fig2}). Of course, increasing $P_{L(R)}$
also allows one to increase the magnitude of the torque (compare
Fig.~\ref{Fig5},right with Fig.~\ref{Fig2}). However, a comparison between the
left and right panels of Figure \ref{Fig5} illustrates that the effects of a
finite $P_{L(R)}$ and a finite SDIPS on the torque are qualitatively
different, since the variations of $A_{20}^{x}$, $A_{20}^{y}$ and $A_{23}^{y}$
with $\delta$ are different in these two panels. Remarkably, in the left
panel, $A_{20}^{x}$ remains positive for any value of $\delta$ whereas it
changes sign with $\delta$ in the right panel.



\section{Measurability of the torque in a colinear geometry\label{col}}

This section discusses the measurability of the spin-torque felt by $L$ in a
simple colinear configuration. I assume that $\overrightarrow{M_{R}}=M_{R}%
\vec{z}^{\prime}$ is fixed along a direction $\vec{z}^{\prime}$ which
corresponds to the anisotropy axis of contact $L$. When one starts from an
initial state $\overrightarrow{M_{L}}=\pm M_{L}\overrightarrow{z}^{\prime}$,
i.e. $\theta=0/\pi$ or $\overrightarrow{z}=\pm\vec{z}^{\prime}$, the
spin-torque vanishes. However, Slonczewski has shown that it is possible to
observe the spin-torque effect by studying the hysteretic switching of
$\overrightarrow{M_{L}}$ between $\theta=0$ and $\theta=\pi$, when a ramping
magnetic field $\overrightarrow{H}$ is applied colinearly to $\overrightarrow
{z}^{\prime}$. Indeed, since $\theta$ must pass continuously between $0$ and
$\pi$ during the switching process, the torques can modify the critical
switching fields. For simplicity, I consider a parameters range where the
torque is approximately sinusoidal, i.e. $T_{m}^{x(y)}\propto\sin[\theta]$. In
the framework of a Landau-Gilbert equation (see appendix), the torque produces
an asymmetry $\Delta H_{sw}=(H_{sw+}+H_{sw-})/2$ of the switching fields
$H_{sw+}$ and $H_{sw-}$ obtained for increasing and decreasing fields, which
can be expressed as
\begin{equation}
\Delta H_{sw}=\left(  \alpha^{-1}\left.  T_{m}^{x}\right\vert _{\theta=\pi
/2}-\left.  T_{m}^{y}\right\vert _{\theta=\pi/2}\right)  /\mu_{0}M_{L}
\label{deltaH}%
\end{equation}
Equation (\ref{deltaH}) involves a Gilbert damping dimensionless constant
$\alpha$, which characterizes the damping of the left magnetization. I will
use below the value $\alpha=0.045$ which has been measured for
nickel\cite{alphaNickel}. The constant $\alpha$ increases the effect of the
in-plane torque $T_{m}^{x}$ with respect to that of the out-of-plane torque
$T_{m}^{y}$.

To motivate experiments, it is interesting to discuss whether the variations
of $T_{m}^{x(y)}$ with the bias or gate voltages are observable through
$\Delta H_{sw}$. Since $\Delta H_{sw}$ scales with the inverse of the
magnetization $M_{L}$, the magnitude of $M_{L}$ must not be too large. I will
assume that the ferromagnet $L$ corresponds to a small magnetic domain in
direct contact with the NC. This domain can belong itself to a larger
ferromagnetic contact, as observed for instance in spin-torque experiments
realized with quantum point contacts\cite{Chen}. Here, I consider a nickel
cubic domain with a side of $2.2$~\textrm{nm}. In spite of this small size, I
disregard Coulomb blockade effects or size-quantization effects inside the
domain\cite{WaintalNanomagnet} since it is assumed to belong to a larger
ferromagnetic contact. The Ni domain encloses about 1000 atoms\cite{Nickel},
and has thus a magnetization $M_{L}=600\mu_{B}$, with $\mu_{B}$ the Bohr
magneton. The transverse spin-dephasing length in Ni is of the order of a
nanometer, as revealed by the superconducting proximity effect observed in
this material\cite{Petrashov,Robinson,Blum,Shelukhin}. Therefore, the
transverse spin-dephasing hypothesis used in this paper seems relevant.
Besides, since the NC is in the few-channels transport regime, I assume that
conduction through the device is limited by the NC/nanodomain contact. In
these conditions, it is reasonable to treat the nanodomain as an electronic
reservoir in equilibrium with $V_{b}^{L}$.

\begin{figure}[ptb]
\includegraphics[width=1.\linewidth]{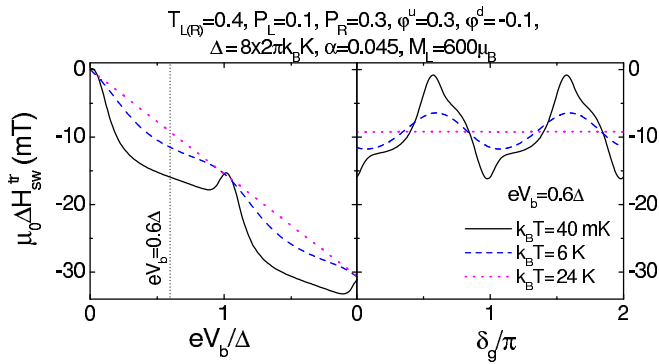}\caption{Spin-torque-induced
asymmetry $\Delta H_{sw}^{tr}$ of the switching fields, in the case where a
finite bias voltage $V_{b}$ is applied to the left reservoir of the spin
valve. The left and right panel show the variations of $\Delta H_{sw}^{tr}$
with $V_{b}$ and the gate-controlled phase $\delta_{g}$, respectively. The
curves are shown for various temperatures. It is assumed that the role of $L$
is played by a nickel cubic domain with a side of $2.2$~\textrm{nm}, which
corresponds to $M_{L}\simeq600\mu_{B}$.}%
\label{Fig3}%
\end{figure}

I use a quadratic band model for the NC, which yields, after a linearization
around the Fermi energy, $\delta=\delta_{g}+(E-E_{F})\pi/\Delta$ \ with
$\delta_{g}=k_{F}\ell+(e\eta V_{g}\pi/\Delta)$ the phase acquired by an
electron with energy $E_{F}$ along the NC. The phase $\delta_{g}$ can be tuned
with $V_{g}$, through a transduction coefficient $\eta$. The parameter
$\Delta=hv_{F}/2\ell$ corresponds to the orbital level spacing inside the NC.
Both thermal regimes $\Delta>k_{B}T$ and $\Delta<k_{B}T$ can be reached in
practice. I assume that the finite bias voltage $V_{b}$ is applied to the left
reservoir ($V_{b}^{L}=V_{b}$ and $V_{b}^{R}=0$). I first discuss the
contribution $\Delta H_{sw}^{tr}$ of $\vec{T}_{m}^{tr}$ to $\Delta H_{sw}$. In
practice, this quantity can be determined by measuring the difference between
the switching fields for $V_{b}=0$ and $V_{b}$ finite. Figure \ref{Fig3} shows
$\Delta H_{sw}^{tr}$ for a given set of interface parameters, and a realistic
value for $\Delta$\cite{delta}. Due to the value of $\alpha$ used, $\Delta
H_{sw}^{tr}$ is dominated by the $T_{m}^{x}$ term ($T_{m}^{x}$ and $T_{m}^{y}$
have comparable amplitudes for the parameters of Fig.~\ref{Fig3}). The left
panel of Fig.\ref{Fig3} shows the dependence of $\Delta H_{sw}^{tr}$ on
$V_{b}$. At low temperatures (i.e. temperatures smaller than the scales of
variation of $A_{20}^{x}$, $A_{20}^{y}$ and $A_{23}^{y}$ with energy), this
dependence is non-linear, due to resonances occurring inside the NC. Besides,
the right panel of Fig.\ref{Fig3} shows that $\Delta H_{sw}^{tr}$ oscillates
with $\delta_{g}$. The amplitude of these oscillations is about $15$%
~\textrm{mT} for the parameters used. In practice, this effect should be
measurable if the switchings of the left magnetic domain are sufficiently
sharp, like observed with single wall carbon nanotubes contacted with
ferromagnets \cite{CPF,Aurich,ote}. Note that for being conservative, I have
used a relatively small SDIPS and small polarizations $P_{L(R)}$ for
estimating $\Delta H_{sw}^{tr}$. In principle, $\Delta H_{sw}^{tr}$ can be
increased significantly by breaking these restrictions. At larger
temperatures, $\Delta H_{sw}^{tr}$ increases linearly with $V_{b}$ and $\Delta
H_{sw}^{tr}$ does not oscillate anymore with $\delta_{g}$.

I now discuss briefly the contribution $\Delta H_{sw}^{eq}$ of $\vec{T}%
_{m}^{eq}$ to $\Delta H_{sw}$. With the parameters of Fig.~\ref{Fig3},right,
at low temperatures, the oscillations of $\Delta H_{sw}^{eq}$ with $\delta
_{g}$ have an amplitude of about $0.2$~\textrm{mT} (not shown), thus $\Delta
H_{sw}\simeq\Delta H_{sw}^{tr}$. Therefore, the gate-induced variations of the
equilibrium torque component in a NC-based spin-valve are probably difficult
to measure in practice.

Note that in Fig. \ref{Fig3}, right, one has $V_{b}\simeq2.6$~\textrm{mV}. In
these conditions, the coherent scattering approach of this paper is relevant.
In order to obtain a current-induced reversal of $\overrightarrow{M_{L}}$ with
no external magnetic field, stronger bias voltages are necessary. In the
latter case, one may have to take into account heating and decoherence effects
inside the NC.

\section{Conclusion}

In this paper, I have discussed the spin-torque effect in a spin-valve made
out of two ferromagnetic leads connected through a coherent NC. I have assumed
that the NC has a single channel near the Fermi energy of the leads. In this
case, the spin-torque effect presents many qualitative differences with
respect to the case of a multichannel disordered spin-valve. I have discussed
the SDIPS-induced interface exchange field, the RKKY-like interlayer exchange
coupling, and the Slonczewski spin-filtering which occur in this device. The
contributions of these three effects to the spin-torque can generally not be
disentangled, due to interference effects occurring inside the NC. One
interesting specificity of a NC-based spin-valve is that the spin-torque can
be modulated with the NC gate voltage. In principle, this modulation can be
observed experimentally, by studying the hysteretic behavior of the spin-valve
with a magnetic field, for instance. This requires to assume that the torque
affects a ferromagnetic nano-domain in direct contact with the NC, and
belonging, for instance, to a wider lithographically defined ferromagnetic contact.

In relation with this work, it is interesting to point out that there also
exists great qualitative differences between the few-channels coherent case
and the multichannel disordered case in the context of non-local spin
transport in a conductor connected to four contacts with colinear
magnetizations\cite{CPF,theorieCPF}.

\textit{I acknowledge discussions with A.-D.~Crisan, T.~Kontos, A.~Thiaville, and X.~Waintal. This work was financially supported by the ANR
under Contract HYFONT No.~09-NANO-002.}

\section{Appendix}

This appendix discusses the dynamics of $\overrightarrow{M_{L}}$ in the
colinear configuration defined in section \ref{col}. The modulus $M_{L}$ of
the magnetization $\overrightarrow{M_{L}}=M_{L}\overrightarrow{m_{L}}$ is
assumed to be constant ($\left\vert \overrightarrow{m_{L}}\right\vert =1$). To
model the dynamics of $\overrightarrow{m_{L}}$, one can use a
Landau-Lifshitz-Gilbert equation\cite{Slonczewski}%
\begin{align}
\frac{d\overrightarrow{m_{L}}}{dt}  &  =-\gamma_{0}\overrightarrow{m_{L}%
}\wedge H\overrightarrow{z}^{\prime}-\gamma_{0}H_{u}(\overrightarrow
{z}.\overrightarrow{m_{L}})\overrightarrow{m_{L}}\wedge\overrightarrow
{z}^{\prime}\nonumber\\
&  +\tau_{//}\overrightarrow{m_{L}}\wedge(\overrightarrow{m_{R}}%
\wedge\overrightarrow{m_{L}})+\tau_{\perp}\overrightarrow{m_{R}}%
\wedge\overrightarrow{m_{L}}\nonumber\\
&  +\alpha\overrightarrow{m_{L}}\wedge\frac{d\overrightarrow{m_{L}}}{dt}
\label{LLG}%
\end{align}
with $\gamma_{0}=-\mu_{0}\gamma>0$, $\gamma\simeq-e/m_{e}$ the gyromagnetic
ratio of electrons and $\mu_{0}$ the vacuum permeability. The uniaxial
anisotropy field of the left ferromagnet is noted $H_{u}\overrightarrow
{z}^{\prime}$. The torque components $T_{m}^{x}$ and $T_{m}^{y}$ occur through
$\tau_{//}=T_{m}^{x}\gamma/M_{L}\sin(\theta)$ and $\tau_{\perp}=-T_{m}%
^{y}\gamma/M_{L}\sin(\theta)$. The Gilbert damping term is proportional to the
dimensionless constant $\alpha$, and usually fulfills $\alpha\ll1$. One can
look for solutions of Eq. (\ref{LLG}) with the form $\overrightarrow{m_{L}%
}=(\sin(\theta)\cos(\omega t),\sin(\theta)\sin(\omega t),\cos(\theta))$ in the
fixed referential $\{x^{\prime},y^{\prime},z^{\prime}\}$. Following
Ref.~\onlinecite{Slonczewski}, one can assume $\omega\gg d\theta/dt$, i.e. the
precession of $\overrightarrow{M_{L}}$ around the $\overrightarrow{z}^{\prime
}$ axis is much faster than its relaxation towards $\pm\overrightarrow{z}$.
This gives $\omega=H\gamma_{0}+\tau_{\perp}+H_{u}\gamma_{0}\cos(\theta)$ and%
\begin{equation}
\frac{d\theta}{dt}=-(\tau_{//}+\alpha\tau_{\perp}+\alpha\gamma_{0}%
H+H_{u}\alpha\gamma_{0}\cos(\theta))\sin(\theta)=F(\theta) \label{oo}%
\end{equation}
Equation (\ref{oo}) corresponds to the dynamics of a fictitious massless
damped particle in an effective potential $U(\theta)$ such that $F(\theta
)=-\partial U(\theta)/\partial\theta$. From Eq. (\ref{oo}), the shape of the
barrier separating the positions $\theta=0$ and $\theta=\pi$ depend on the
torques. Here, I assume that $T_{m}^{x}$ and $T_{m}^{y}$ are approximately
sinusoidal, so that $\tau_{\perp}$ and $\tau_{//}$ can be treated as
constants. In this case, $\overrightarrow{m_{L}}$ can switch from
$\pm\overrightarrow{z}$ to $\mp\overrightarrow{z}$ if $H$ decreases/increases
until it reaches the value $H_{sw\pm}=\mp H_{u}-((\tau_{//}/\alpha
)+\tau_{\perp})/\gamma_{0}$.


\begin{thebibliography}{99}
\bibitem {note}This approach differs from the scattering approach of Refs.
~\onlinecite{Waintal,Slonczewski}, where the ferromagnet $L$ is not treated as
a reservoir and the torque is calculated as the difference of the
spin-currents at the left and right sides of $L$. However, in the limit of a
strong transverse spin dephasing, such complications are not
necessary\cite{Stiles,Brataas1,Brataas2}.
\bibitem {Nickel}Nickel has a face-centered cubic structure with a lattice
parameter of about 0.35~nm.
\bibitem {ote}It has been checked from MFM characterizations that the contacts
used in Ref. \onlinecite{CPF} presented many small magnetic domains. This did
not prevent the appearance of sharp switchings.
\end{thebibliography}
\end{document}